\documentclass[aps,prl,superscriptaddress,floatfix,showpacs,a4paper]{revtex4}

\usepackage{graphicx,graphics,epsfig}
\usepackage{dcolumn}
\usepackage{bm}
\usepackage{amsmath}
\usepackage{verbatim}
\usepackage{color}
\usepackage{subfigure}
\usepackage{times,natbib}
\usepackage{amsmath,amsfonts,amssymb,graphics,graphicx,epsfig,color,times,natbib}

\begin{document}
\title{Testing Leggett's Inequality Using Aharonov-Casher Effect}

\author{Hong-Yi Su}
\affiliation{Theoretical Physics Division, Chern Institute of
Mathematics, Nankai University, Tianjin 300071, People's Republic of
China}

\author{Jing-Ling Chen\footnote{Correspondence and requests for materials should be addressed
to J.L.C. (cqtchenj@nus.edu.sg).}}
\affiliation{Theoretical Physics Division, Chern Institute of
Mathematics, Nankai University, Tianjin 300071, People's Republic of
China} \affiliation{Centre for Quantum Technologies, National
University of Singapore, 3 Science Drive 2, Singapore 117543}

\author{Chunfeng Wu}
\affiliation{Centre for Quantum Technologies, National University of
Singapore, 3 Science Drive 2, Singapore 117543} \affiliation{Pillar
of Engineering Product Development, Singapore University of
Technology and Design, 20 Dover Drive, Singapore 138682}

\author{Dong-Ling Deng}
\affiliation{Department of Physics and Michigan Center for
Theoretical Physics, University of Michigan, Ann Arbor, Michigan
48109, USA}

\author{C. H. Oh}
\affiliation{Centre for Quantum Technologies, National University of
Singapore, 3 Science Drive 2, Singapore 117543}
\affiliation{Department of Physics, National University of
Singapore, 2 Science Drive 3, Singapore 117542}

\date{\today}

\begin{abstract}
\textbf{Bell's inequality is established based on local realism. The
violation of Bell's inequality by quantum mechanics implies either
locality or realism or both are untenable. Leggett's inequality is
derived based on nonlocal realism. The violation of Leggett's
inequality implies that quantum mechanics is neither local realistic
nor nonlocal realistic. The incompatibility of nonlocal realism and
quantum mechanics has been currently confirmed by photon
experiments. In our work, we propose to test Leggett's inequality
using the Aharonov-Casher effect. In our scheme, four entangled
particles emitted from two sources manifest a two-qubit-typed
correlation that may result in the violation of the Leggett
inequality, while satisfying the no-signaling condition for
spacelike separation. Our scheme is tolerant to some local
inaccuracies due to the topological nature of the Aharonov-Casher
phase. The experimental implementation of our scheme can be possibly
realized by a calcium atomic polarization interferometer experiment.}
\end{abstract}

\pacs{03.65.Ta, 03.65.Ud, 03.67.-a}

\maketitle

Bell's inequality~\cite{Bell,J.Clauser} imposes bounds on
correlations of different parties of multipartite systems based on
local realism. However, the violation of Bell's inequality by
quantum mechanics implies either locality or realism or both are
untenable. In the debate of incompatibility between quantum
mechanics and any local realistic hidden variable theory,
experiments~\cite{experiment1,loophole1,loophole2,loophole3,loophole4,loophole5,loophole6}
have supported quantum mechanics. Although for some time there
existed loopholes of
locality~\cite{loophole1,loophole2,loophole4,loophole5} and
detection~\cite{loophole3,loophole6}, they have been almost closed.
The invalidity of local realism is a reasonably established fact. In
2003, Leggett~\cite{Leggett} derived a class of inequalities based
on nonlocal realism. He assumed that the state of a subsystem has
been predetermined by some variable $\lambda$ even before the
measurement, and that the joint probabilities consist of a mixture
of correlations that cannot be separable. Since only states of
subsystems have been predetermined, the whole system may be
nonlocal. After the pioneer work, the incompatibility
of nonlocal realism and quantum mechanics was experimentally confirmed 
~\cite{LeggettEx1,LeggettEx2,LeggettEx3,2008BRANCIARD}. It was shown that quantum mechanics is neither
local realistic nor nonlocal realistic.

Topological property of physical systems has given rise to many applications
ranging from quantum field theory to quantum information science. An
example is the Aharonov-Bohm (AB)
effect~\cite{ABeffect}, in which a moving charge has its phase
shifted in the presence of a confined magnetic field, though apparently it feels
no net force. In 1984, Aharonov and Casher~\cite{ACeffect} predicted
a dual of the AB effect. In the Aharonov-Casher (AC) effect, the
role of charge and magnetic flux is exchanged, i.e., when a neutral
particle with magnetic moment moves around an impenetrable line
charge, it also acquires some phase shifts. The AC effect is
traditionally understood as a nonlocal and topological effect in
which a particle with magnetic moment acquires shifted phase when
moving in a topologically nontrivial region.
In 1998 a scheme involving the AC effect to
test local realism was proposed by Pati~\cite{testBIviaAC2}, and the
violation of Bell's inequality indicates the nonlocality of the four-particle
entangled state. 

In this work, we advance the study of nonlocality of the AC effect
and present a scheme to test Leggett's inequality by resorting to
the AC effect. Due to the topological nature of the AC effect, our
scheme is robust against some local inaccuracies. We shall test the
two-qubit Leggett inequality in a physical system consisting of four
neutral particles with magnetic moments, whose initial state is a
product of the singlet state of pair $(1,2)$ and the triplet state
of pair $(3,4)$. Pseudo-Pauli matrices are introduced such that one
may view an entangled spin pair like pair $(1,4)$ or pair $(2,3)$ as
a ``single qubit", and hence four particles as a total ``two-qubit"
system. The influence of the AC effect on an entangled spin pair is
found to be equivalent to a rotation in terms of the pseudo-Pauli
operators. Moreover, based on the final state of the four particles,
we focus on some specific joint probabilities satisfying the
no-signaling condition and obtain a two-qubit-type correlation
function that may violate the Leggett inequality. We also present
some discussion on the implementation of our scheme in a calcium
atomic polarization interferometer experiment.

\section{Results}

\textbf{Testing Leggett's Inequality using the AC Effect.} In a nonlocal hidden
variable model, one assumes that the joint probability for a
bipartite system consists of statistical mixture of simpler
correlations:
\begin{eqnarray}
P(\alpha,\beta|\vec{a},\vec{b})=\int
d\lambda\rho(\lambda)P_{\lambda}(\alpha,\beta|\vec{a},\vec{b}),
\end{eqnarray}
where $\lambda$ is a set of hidden variables determining the system,
$\rho(\lambda)$ distribution of $\lambda$; $\alpha,\beta$ are
measurement outcomes, and $\vec{a},\vec{b}$ measurement settings for
two subsystems, respectively. An extra requirement is that
$P_{\lambda}$ satisfies the no-signaling condition, i.e.,
$\sum_{\beta}P_{\lambda}(\alpha,\beta|\vec{a},\vec{b})=\sum_{\beta}P_{\lambda}(\alpha,\beta|\vec{a},\vec{b'})$
and
$\sum_{\alpha}P_{\lambda}(\alpha,\beta|\vec{a},\vec{b})=\sum_{\alpha}P_{\lambda}(\alpha,\beta|\vec{a'},\vec{b})$.
Follow Branciard {\it et al.}'s derivation of the Leggett
inequality~\cite{2008BRANCIARD}, one can define the correlations for
a two-qubit system as $P_{\lambda}(\alpha,\beta|\vec{a},\vec{b})
=\frac{1}{4}(1+\alpha M^A_{\lambda}(\vec{a},\vec{b})+\beta
M^B_{\lambda}(\vec{a},\vec{b})+\alpha\beta
C_{\lambda}(\vec{a},\vec{b}))$.
Here $M^A_{\lambda}(\vec{a},\vec{b})=\sum_{\alpha,\beta}\alpha
P_{\lambda}(\alpha,\beta|\vec{a},\vec{b})$,
$M^B_{\lambda}(\vec{a},\vec{b})=\sum_{\alpha,\beta}\beta
P_{\lambda}(\alpha,\beta|\vec{a},\vec{b})$,
$C_{\lambda}(\vec{a},\vec{b})=\sum_{\alpha,\beta}\alpha\beta
P_{\lambda}(\alpha,\beta|\vec{a},\vec{b})$, and $\alpha,\beta=\pm
1$. $M^A_{\lambda}$ and $M^B$ are expectation values (or marginals)
at each respective measuring location.
According to no-signaling condition, the marginals $M^A_{\lambda}$
and $M^B_{\lambda}$ can be locally described by their respective
choices of measurement, i.e.,
$M^A_{\lambda}(\vec{a},\vec{b})=M^A_{\lambda}(\vec{a})$ and
$M^B_{\lambda}(\vec{a},\vec{b})=M^B_{\lambda}(\vec{b})$. Leggett
assumed that each subsystem can be described by a pure quantum
state, then for the two-qubit case, each hidden variable determines
a product state
$\lambda\rightarrow|\vec{u}\rangle\otimes|\vec{v}\rangle$, where
$\vec{u},\vec{v}$ are vectors on the Poincar\'{e} sphere. We have
$M^A_{\lambda}(\vec{a})=\langle
\vec{\sigma}\cdot\vec{a}\rangle_{\lambda}=\vec{u}_{\lambda}\cdot
\vec{a}$, $M^B_{\lambda}(\vec{b})=\langle
\vec{\sigma}\cdot\vec{b}\rangle_{\lambda}=\vec{v}_{\lambda}\cdot
\vec{b}$, where $\vec{\sigma}=(\sigma^x, \sigma^y, \sigma^z)$ is the
usual Pauli matrix vector. Since no further assumption of the
bipartite correlation function is made, generally speaking,
$C_{\lambda}(\vec{a},\vec{b})\neq
M^A_{\lambda}(\vec{a})M^B_{\lambda}(\vec{b})$. The Leggett
inequality is of the following form~\cite{2008BRANCIARD}
\begin{eqnarray}
|C(\vec{a},\vec{b})\pm C(\vec{a},\vec{b'})|\leq 2-\int d\lambda
\rho(\lambda)|M^B_{\lambda}(\vec{b})\mp M^B_{\lambda}(\vec{b'})|,
\end{eqnarray}
where $\vec{a}$ describes the measurement setting for Alice,
$\vec{b}, \vec{b}'$ are the two measurement settings for Bob.
Consider three settings $\vec{a}_i (i=1,2,3)$ for Alice and six
settings $\vec{b}_i, \vec{b}'_i$ $(i=1,2,3)$ for Bob as given in
Ref. \cite{2008BRANCIARD}, where
$\vec{b}_i-\vec{b}'_i=2\sin(\varphi/2)\vec{e}_i$ with $\vec{e}_i
(i=1,2,3)$ being an orthogonal basis, then one arrives at the
Leggett inequality as
\begin{eqnarray}\label{Leggett0}
\frac{1}{3}\sum_{i=1}^3(|C(\vec{a}_i,\vec{b}_i)+
C(\vec{a}_i,\vec{b'}_i)|) \leq
2-\frac{2}{3}\biggr|\sin{\frac{\varphi}{2}}\biggr|.
\end{eqnarray}
For the singlet state of two qubits, the quantum correlation
function reads $C(\vec{a},\vec{b})=\int d\lambda
\rho(\lambda)C_{\lambda}(\vec{a},\vec{b})=\langle
\vec{\sigma}\cdot{\vec{a}}\otimes\vec{\sigma}\cdot{\vec{b}}\rangle=-\vec{a}\cdot\vec{b}$.
Then the Leggett inequality reduces to a simpler form:
\begin{eqnarray}
2\biggr|\cos{\frac{\varphi}{2}}\biggr|+\frac{2}{3}\biggr|\sin{\frac{\varphi}{2}}\biggr|\leq2.\label{Leggett}
\end{eqnarray}

Let us consider a system of four neutral spin-1/2 particles with
magnetic moments in the presence of a line charge. In the AC
configuration, the particles are moving in $xy$-plane and the line
charge oriented along the third axis (the $z$-axis). The motion of
the particles is influenced by the electric field of line charge.
Each particle will acquire a phase when moving along the plane,
\begin{eqnarray}\label{ACtransform}
&&|\uparrow\rangle_j\rightarrow e^{i\int_{\ell_j}
(\vec{E}\times\vec{\mu}_j)\cdot
d\vec{r}}|\uparrow\rangle_j=e^{i\frac{\varphi_j}{2}}|\uparrow\rangle_j,\nonumber\\
&&|\downarrow\rangle_j\rightarrow e^{-i\int_{\ell_j}
(\vec{E}\times\vec{\mu}_j)\cdot
d\vec{r}}|\downarrow\rangle_j=e^{-i\frac{\varphi_j}{2}}|\downarrow\rangle_j,
\end{eqnarray}
where $|\uparrow\rangle_j$, $|\downarrow\rangle_j$ describe quantum
states with spin up and spin down for the $j$-th particle, $\vec{E}$ is electric field intensity,
$\varphi_j=\int_{\ell_j} (\vec{E}\times\vec{\mu}_j)\cdot d\vec{r}$ is
the measurable phase accumulated during the evolution, and
$\vec{\mu}_j$ is the magnetic moment for the $j$-th particle.

Denote the state for two spin-1/2 particles by $|S,M\rangle$ with
total spin $S$ and magnetic quantum number $M$, then the singlet
state and the triplet state with $M=0$ are given by
\begin{eqnarray}\label{singlet}
&&|0,0\rangle=(|\uparrow\downarrow\rangle-|\downarrow\uparrow\rangle)/{\sqrt{2}},\nonumber\\
&&|1,0\rangle=(|\uparrow\downarrow\rangle+|\downarrow\uparrow\rangle)/{\sqrt{2}},
\end{eqnarray}
Under the transformation (\ref{ACtransform}), the initial singlet
state $|0,0\rangle_{mn}$ and triplet state $|1,0\rangle_{mn}$ of
particles $m$ and $n$ become
\begin{eqnarray}\label{transform1}
&&|0,0\rangle_{mn}\rightarrow\cos{\frac{\varphi_m-\varphi_n}{2}}|0,0\rangle_{mn}+i\sin{\frac{\varphi_m-\varphi_n}{2}}|1,0\rangle_{mn},\nonumber\\
&&|1,0\rangle_{mn}\rightarrow
i\sin{\frac{\varphi_m-\varphi_n}{2}}|0,0\rangle_{mn}+\cos{\frac{\varphi_m-\varphi_n}{2}}|1,0\rangle_{mn},
\end{eqnarray}
namely, the states $|0,0\rangle_{mn}$ and $|1,0\rangle_{mn}$ evolve
to the quantum states that are linear superpositions of themselves.
This is a very notable feature of the AC effect influencing an
entangled spin pair \cite{testBIviaAC2,testBIviaAC1}. Equation
(\ref{transform1}) implies that $\{|0,0\rangle, |1,0\rangle\}$ may
span a subspace, and in turn one may treat the spin pair as a
``single qubit". To make this point explicit, let us abbreviate
\begin{eqnarray}\label{definition1}
&&|\overline{0}\rangle\equiv|0,0\rangle,\;\;\;|\overline{1}\rangle\equiv|1,0\rangle,
\end{eqnarray}
then Eq. (\ref{transform1}) can be recast as
\begin{eqnarray}\label{transform2}
\left(
\begin{array}{l}|\overline{0}\rangle_{mn}\\
|\overline{1}\rangle_{mn} \end{array} \right) \rightarrow \left(
\begin{array}{cc}\cos{\frac{\varphi_m-\varphi_n}{2}}& i\sin{\frac{\varphi_m-\varphi_n}{2}}\\
i\sin{\frac{\varphi_m-\varphi_n}{2}}&\cos{\frac{\varphi_m-\varphi_n}{2}}
\end{array} \right)\left(
\begin{array}{l}|\overline{0}\rangle_{mn}\\
|\overline{1}\rangle_{mn} \end{array} \right).
\end{eqnarray}
Moreover, one defines the following pseudo-Pauli matrices as
${\Sigma}^x=|\overline{0}\rangle\langle
\overline{1}|+|\overline{1}\rangle\langle \overline{0}|$,
${\Sigma}^y=-i(|\overline{0}\rangle\langle
\overline{1}|-|\overline{1}\rangle\langle \overline{0}|)$,
${\Sigma}^z=|\overline{0}\rangle\langle
\overline{0}|-|\overline{1}\rangle\langle \overline{1}|,$
which share similar properties as the usual Pauli matrices, then Eq.
(\ref{transform2}) is nothing but a rotation
\begin{eqnarray}\label{Rx}
\mathcal{R}^x_{mn}(\varphi_m-\varphi_n)=e^{i(\varphi_m-\varphi_n){\Sigma}^x_{mn}/2}
\end{eqnarray}
along $x$-axis on the basis $\{|\overline{0}\rangle,
|\overline{1}\rangle\}$ of the ``single qubit".

Our scheme for testing the Leggett inequality by experiment involves
two pairs of entangled spin-1/2 particles. Similar to
Refs.~\cite{testBIviaAC1,testBIviaAC2}, we prepare the four
particles entangled in two pairs (1,2) and (3,4) initially, and
finally perform some proper projective measurements on particle
pairs $(1,4)$ and $(2,3)$ to obtain the correlation function. Assume
initially that particles 1 and 2 are emitted from a source $O_{12}$
with total spin $S_{12}=0$ and magnetic moment $M_{12}=0$;
similarly, particles 3 and 4 are emitted from a source $O_{34}$ with
$S_{34}=1$ and $M_{34}=0$. Namely, the initial state reads
$|\Psi_{\rm i}\rangle=|0,0\rangle_{12}\otimes|1,0\rangle_{34}
=\frac{1}{2}(|\uparrow\downarrow\downarrow\uparrow\rangle-
|\downarrow\uparrow\uparrow\downarrow\rangle+|\uparrow\uparrow\downarrow\downarrow\rangle
-|\downarrow\downarrow\uparrow\uparrow\rangle)_{1423}$, in the last
step of which we have rearranged the particles in the order of
``1423". Actually, $|\Psi_{\rm i}\rangle$ can be rewritten as
\begin{eqnarray}\label{initial1a}
|\Psi_{\rm
i}\rangle=\frac{1}{2}(|\overline{0}\rangle_{14}|\overline{1}\rangle_{23}-|\overline{1}\rangle_{14}\overline{0}\rangle_{23})
+\frac{1}{2}(|\uparrow\uparrow\downarrow\downarrow\rangle-|\downarrow\downarrow\uparrow\uparrow\rangle)_{1423}.
\end{eqnarray}
However, the last two terms of Eq. (\ref{initial1a}) will vanish
when they are acted by any operator defined in the subspace
$\overline{\mathcal {H}}=\{|\overline{0}\rangle_{14},
|\overline{1}\rangle_{14}\}\otimes \{|\overline{0}\rangle_{23},
|\overline{1}\rangle_{23}\}$. Here we retain them for normalization.
In fact, the initial state can be understood as a ``singlet state"
$|\Psi_{\rm i}\rangle \propto
\frac{1}{\sqrt{2}}(|\overline{0}\rangle_{14}|\overline{1}\rangle_{23}-|\overline{1}\rangle_{14}\overline{0}\rangle_{23})$
of ``two-qubit" without any confusion.

\begin{figure}[tbp]
\includegraphics[width=120mm]{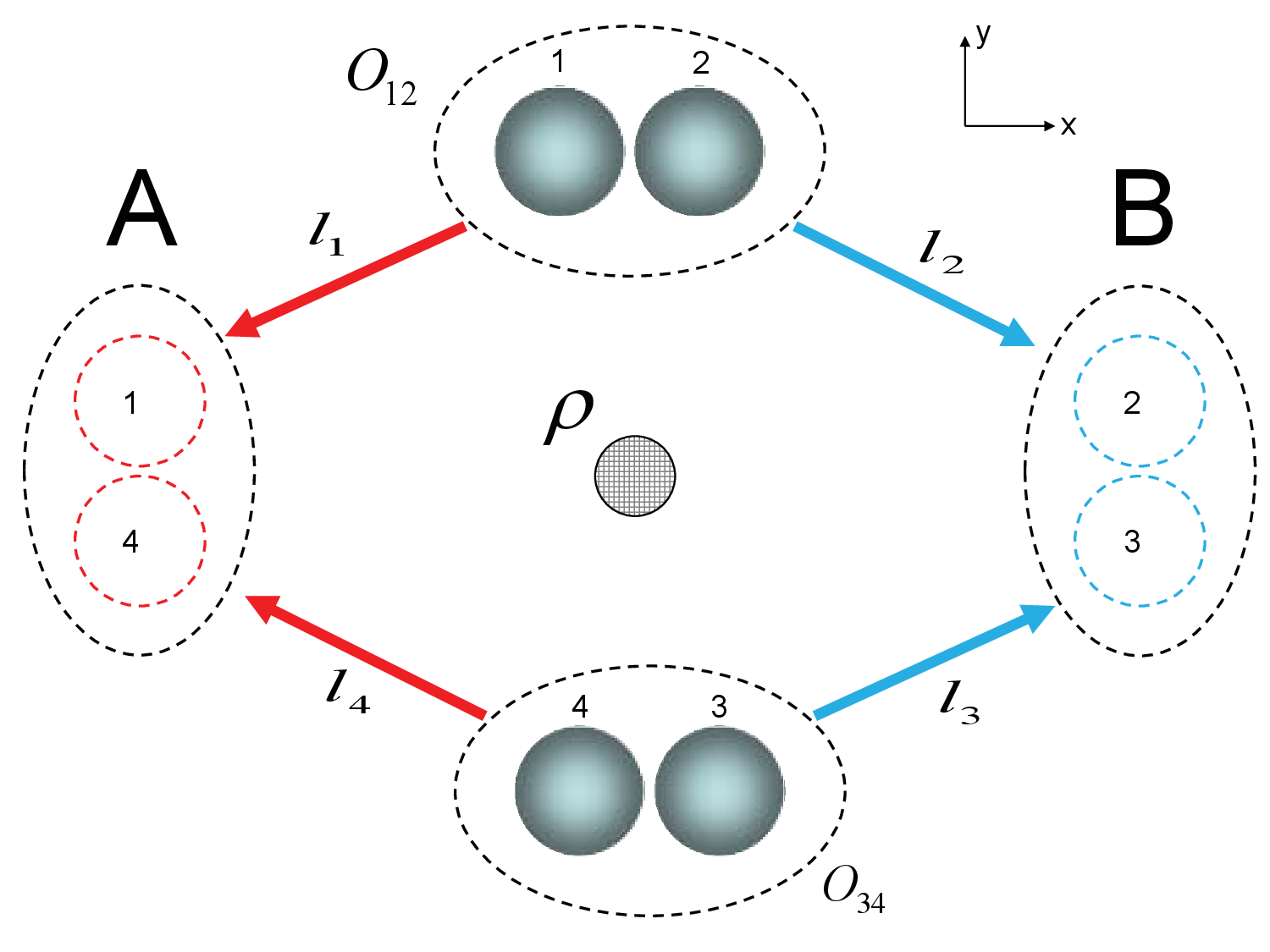}\\
\caption{\textbf{A schematic illustration of experiment proposal.}
We let the two sources be located at points $O_{12}$ and $O_{34}$ on
the $xy$-plane respectively, and invoke an impenetrable line charge
(with charge density $\rho$) oriented along the $z$-axis. After the
four particles are emitted from the two sources, we then move
particle 1 from location $O_{12}$ to location $A$ along path
$\ell_1$, and move particle 4 from location $O_{34}$ to meet
particle 1 at location $A$ along path $\ell_4$. The motion of the
particles are influenced by the electric field of line charge as
shown in Eq. (\ref{ACtransform}) and accordingly the corresponding
AC phase shifts are $\varphi_1$ and $\varphi_4$ for particles 1 and
4 respectively. Similarly, we move particle 2 from location $O_{12}$
to location $B$ along path $\ell_2$, and move particle 3 from
location $O_{34}$ to meet particle 2 at location $B$ along path
$\ell_3$, and the corresponding AC phase shifts are $\varphi_2$ and
$\varphi_3$ for particles 2 and 3 respectively. } \label{fig1}
\end{figure}

Our experiment proposal is demonstrated in Fig.~\ref{fig1}. 
The distance from $A$ to $B$ is
supposed to be large enough so that the measurement of particle pair
(1,4) and that of particle pair (2,3) are space-like, and thus
no-signaling condition is satisfied.
Due to Eqs. (\ref{transform2}) and (\ref{Rx}), we have the final
state of the four particles as
\begin{eqnarray}
|\Psi_{\rm f}\rangle
=\frac{1}{2}\mathcal{R}^x_A(\varphi_A)\otimes\mathcal{R}^x_B(\varphi_B)
(|\overline{0}\rangle_{14}|\overline{1}\rangle_{23}-|\overline{1}\rangle_{14}\overline{0}\rangle_{23})
+\frac{1}{2}(e^{i\gamma}|\uparrow\uparrow\downarrow\downarrow\rangle_{1423}
-e^{-i\gamma}|\downarrow\downarrow\uparrow\uparrow\rangle_{1423}).\label{final}
\end{eqnarray}
Here $A$ represents ``14" and $B$ represents ``23",
$\gamma=(\varphi_1+\varphi_4-\varphi_2-\varphi_3)/2$, and
$\varphi_A=\varphi_1-\varphi_4$, $\varphi_B=\varphi_2-\varphi_3$ are
relative AC phases for meeting locations A and B acquired by four
particles moving along different paths. It is worth to mention that
AC effect usually concerns a single particle moving around a line
charge, however here none of the moving paths of four particles
encircles the line charge, though the combination of four
corresponding paths actually makes a circle.

Next we perform local projective measurements on two particle pairs
(1,4) and (2,3) along arbitrary directions
$\vec{n}_A=(\sin\xi_A\cos\theta_A,\sin\xi_A\sin\theta_A,\cos\xi_A)$
and
$\vec{n}_B=(\sin\xi_B\cos\theta_B,\sin\xi_B\sin\theta_B,\cos\xi_B)$,
respectively. The projectors are defined as $\hat{\mathcal
{P}}(i,j)=|\overline{i}_{n_A}\overline{j}_{n_B}\rangle\langle
\overline{i}_{n_A}\overline{j}_{n_B}|$, $(i,j=0,1)$, where
\begin{eqnarray}\label{proj1}
|\overline{0}_{\vec{n}}\rangle & = &
(|+\vec{n},-\vec{n}\rangle-|-\vec{n},+\vec{n}\rangle)/{\sqrt{2}},\nonumber\\
|\overline{1}_{\vec{n}}\rangle & = &
(|+\vec{n},-\vec{n}\rangle+|-\vec{n},+\vec{n}\rangle)/{\sqrt{2}},
\end{eqnarray}
which are respectively the singlet state and the triplet state with
$M=0$ written in terms of the following states: $|+\vec{n}\rangle =
\cos{\frac{\xi}{2}} |\uparrow\rangle + \sin{\frac{\xi}{2}}e^{i
\theta} |\downarrow\rangle$, $|-\vec{n}\rangle  =
\sin{\frac{\xi}{2}} |\uparrow\rangle - \cos{\frac{\xi}{2}}e^{i
\theta} |\downarrow\rangle$. Here we choose the vectors $\vec{n}_A$
and $\vec{n}_B$ in the $xy$-plane, i.e., $\xi_A=\xi_B=\pi/2$. Let us
denote $P(i,j)=\langle\Psi_{\rm f}|\hat{\mathcal {P}}(i,j)|\Psi_{\rm
f}\rangle$ as the joint probabilities satisfying the no-signaling
condition, and based on which the correlation function is defined as
\begin{eqnarray}\label{correlation1}
C_{AB}&=&\frac{
\sum_{i,j=0,1}(-1)^{i+j}P(i,j)}{\sum_{i,j=0,1}P(i,j)}.
\end{eqnarray}
After some calculations, we obtain the explicit result of the
correlation function as
\begin{eqnarray}\label{correlation2}
C_{AB}(\vec{a},\vec{b})
&=& -\vec{a}\cdot\vec{b},
\end{eqnarray}
where  $\vec{a}=(\sin\theta_A \cos\varphi_A, \sin\theta_A
\sin\varphi_A, \cos\theta_A)$ and $\vec{b}=(\sin\theta_B
\cos\varphi_B, \sin\theta_B \sin\varphi_B, \cos\theta_B)$ are two
unit three-dimensional vectors. Here the vectors $\vec{a}$ and
$\vec{b}$ (or say $\varphi_A,\varphi_B$, $\theta_A,\theta_B$) are
experimentally controllable: The parameters $\varphi_A,\varphi_B$
(i.e., $\varphi_i,\;i=1,2,3,4$) are the relative AC phase shifts
of the four particles determined by the locations $A, B$ and the paths
$\ell_i$; and the parameters $\theta_A,\theta_B$ come from the
selection of directions in the projective measurements for each
particle pair at $A$ and $B$.
Actually, the correlation function (\ref{correlation2}) is
equivalent to $C_{AB}(\vec{a},\vec{b})=\langle \Psi_{\rm
i}|\vec{\Sigma}\cdot{\vec{a}}\otimes\vec{\Sigma}\cdot{\vec{b}}|\Psi_{\rm
i}\rangle$, which is just similar to that of two usual qubits under
the joint measurement
$\vec{\sigma}\cdot\vec{a}\otimes\vec{\sigma}\cdot\vec{b}$ on the
singlet state. This correspondence also provides a reasonable
explanation on why the AC effect can be used to test
both the Bell-Clauser-Horne-Shimony-Holt (Bell-CHSH)
inequality~\cite{J.Clauser} in Ref.~\cite{testBIviaAC2} and the
Leggett inequality in this work.

Reference \cite{testBIviaAC2} proposed to test the Bell-CHSH inequality
\begin{eqnarray}\label{CHSH}
|C(\vec{a},\vec{b})+C(\vec{a}',\vec{b})+C(\vec{a},\vec{b}')-C(\vec{a}',\vec{b}')|\le
2
\end{eqnarray}
using the AC effect. There are four measurement settings in the
inequality (\ref{CHSH}), i.e., $\vec{a},\vec{b},\vec{a}',\vec{b}'$.
To attain maximal violation of the inequality, it is sufficient to
put the four measurement settings in the same plane, i.e., one may
always choose $\theta_A=\theta_{A'}=\theta_B=\theta_{B'}=\pi/2$ if
the Bell-CHSH inequality is tested. By properly selecting two
locations $A, A'$ for Alice where particle pair (1,4) meets, and two
locations $B, B'$ for Bob where particle pair (2,3) meets, and
adjusting the phase shifts as $\varphi_{A}=0$, $\varphi_{A'}=\pi/2$,
$\varphi_{B}=\pi/4$, $\varphi_{B'}=-\pi/4$, or say $\vec{a}=(1, 0,
0)$, $\vec{a}'=(0, 1, 0)$, $\vec{b}=(1/\sqrt{2}, 1/\sqrt{2}, 0)$,
and $\vec{b}'=(1/\sqrt{2}, -1/\sqrt{2}, 0)$, then the right-hand
side of (\ref{CHSH}) achieves $2\sqrt{2}$ and thus the Bell-CHSH
inequality is maximally violated. The violation of the Bell
inequality rules out local realistic theories from quantum
mechanics.

To test Leggett's inequality (\ref{Leggett0}), we need totally nine
measurement settings, i.e.,
$\vec{a}_1,\vec{a}_2,\vec{a}_3,\vec{b}_1,\vec{b}'_1,\vec{b}_2,\vec{b}'_2,\vec{b}_3,\vec{b}'_3$.
Since $\vec{e}_i (i=1,2,3)$ is an orthogonal basis, the nine
measurement settings cannot lie in the same plane. Properly
select three locations $A_i (i=1,2,3)$ for Alice where particle
pair (1,4) meets, and six locations $B_i/B'_i (i=1,2,3)$ for Bob
where particle pair (2,3) meets (see Fig. 2), and adjust the nine
different paths and nine directions of the projectors such that the
measurement settings are $(\theta_{A_1}, \varphi_{A_1})=(\pi/2, 0)$,
$(\theta_{A_2}, \varphi_{A_2})=(\pi/2, \pi/2)$, $(\theta_{A_3},
\varphi_{A_3})=(0, 0)$, $(\theta_{B_1}, \varphi_{B_1})=(\pi/2,
\varphi/2)$, $(\theta_{B'_1}, \varphi_{B'_1})=(\pi/2, -\varphi/2)$,
$(\theta_{B_2}, \varphi_{B_2})=(\pi/2-\varphi/2, \pi/2)$,
$(\theta_{B'_2}, \varphi_{B'_2})=(\pi/2+\varphi/2, \pi/2)$,
$(\theta_{B_3}, \varphi_{B_3})=(\varphi/2, 0)$, $(\theta_{B'_3},
\varphi_{B'_3})=(\varphi/2, \pi)$, we arrive at the experimental
settings given in Ref.~\cite{2008BRANCIARD}. Based on which the six
correlation functions in Eq. (\ref{Leggett0}) are all equal to
$-\cos(\varphi/2)$, and consequently for $|\varphi|\in (0,
4\tan^{-1}(\frac{1}{3}))$, the Leggett inequality~(\ref{Leggett}) is
violated. The violation of the Leggett inequality implies that
nonlocal realistic theories are not compatible with quantum
mechanics. In the AC experiment, the invalidity of both the Bell
inequality and the Leggett inequality suggests that quantum
mechanics is neither local nor realistic. The result is consistent
with the works in the
literatures~\cite{LeggettEx1,LeggettEx2,LeggettEx3,2008BRANCIARD}
based on the experiment of entangled photons.

\section{Discussion}
Let us make some discussion on the possible implementation of our
scheme in physical systems. One possible system to explore the our
scheme experimentally is a calcium atomic polarization
interferometer as investigated in Ref. \cite{Shinya2002}. Encode two
magnetic substates of the excited state $|^3P_1\rangle$ as
computational basis, $|\uparrow\rangle \equiv|^{3}P_1,+1\rangle$ and
$|\downarrow\rangle\equiv|^{3}P_1,-1\rangle$, the phase difference
between $|\uparrow\rangle$ and $|\downarrow\rangle$ accumulated
during the evolution includes two parts, one is dynamical
phase and the other one is nothing but the AC phase. As we know, the
presence of dynamical phase may destroy the potential robustness of
our scheme since it is sensitive to noise. Fortunately the dynamical part can be canceled out via
interferometer, as shown in Ref. \cite{Shinya2002}, and therefore one
only has the AC phase in the experiment. Due to the topological
property of the AC phase, the experiment offers a promising
fault-tolerant method to test Leggett's inequality. The experimental
achievement in the literature~\cite{Shinya2002} tells us that our
test of Leggett's inequality using the AC effect is possibly
realizable with current techniques in an experiment of a calcium
atomic polarization interferometer.

In summary, we have proposed a scheme to test the two-qubit Leggett
inequality using two entangled spin-1/2 particle pairs emitted from
two sources in the presence of a line charge. Pseudo-Pauli matrices
are introduced such that these four particles can be viewed as a
total ``two-qubit" system. The influence of the AC effect on each
entangled spin pair is found to be equivalent to a rotation in terms
of the pseudo-Pauli operators. Based on the final state of the
physical system, two-qubit-type correlation functions with
controllable parameters can be calculated from joint probabilities
for the measurement of the two particle pairs with $M=0$. The
Leggett inequality is found to be violated, which implies the
invalidity of nonlocal realistic theories. The merit of our scheme
lies at robustness against local inaccuracy, and thus our scheme of testing
the Leggett inequality is tolerant to some local inaccuracies. As is
well known, photon-based experiments often encounter loophole
problems, such as errors in the detectors and detecting systems. The
existence of loopholes may affect the validity of the experiments,
and hence the investigation of loophole-free experiments is a good
alternative.
This makes our scheme totally different from the known
experiments on testing the Leggett inequality in the
literatures~\cite{LeggettEx1,LeggettEx2,LeggettEx3,2008BRANCIARD}.

\vspace{3mm}

\indent{\bf Acknowledgements}

We thank Prof A. J. Leggett for very helpful suggestions and
valuable comments. J.L.C. is supported by National Basic Research
Program (973 Program) of China under Grant No. 2012CB921900, NSF of
China (Grant Nos. 10975075 and 11175089). This work is partly supported by
National Research Foundation and Ministry of Education of Singapore
(Grant No. WBS: R-710-000-008-271).

\vspace{3mm}

{\bf Author contributions}

JLC initiated the idea. JLC, HYS and CW derived the results. JLC,
HYS, CW, DLD, and CHO wrote the main manuscript text. HYS prepared
the figure. All authors reviewed the manuscript.

\vspace{3mm}

{\bf Additional information}

\textbf{Competing financial interests:} The authors declare no
competing financial interests.

\newpage

\begin{figure}[tbp]
\includegraphics[width=150mm]{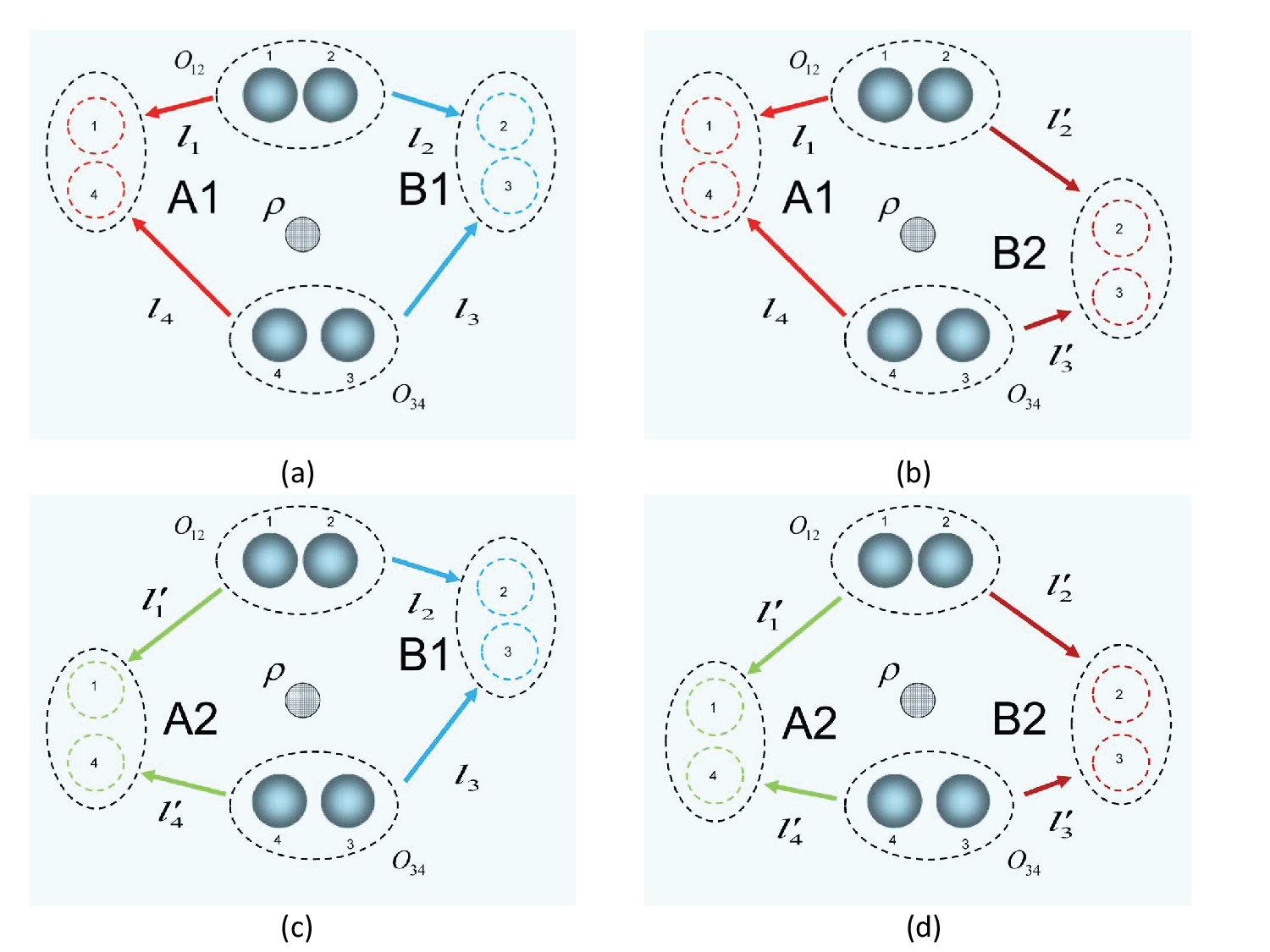}\\
\caption{\textbf{Illustration of different locations and
trajectories in space.} Properly choose three locations $A_i
(i=1,2,3)$ for Alice where particle pair (1,4) meets, and six
locations $B_i/B'_i (i=1,2,3)$ for Bob where particle pair (2,3)
meets, and control the different paths such that we arrive at the
experimental settings given in Ref.~\cite{2008BRANCIARD}. (a)
Illustration of locations $A_1$ and $B_1$, and paths $\ell_1,
\ell_2, \ell_3, \ell_4$; (b) Illustration of locations $A_1$ and
$B_2$, and paths $\ell_1, \ell'_2, \ell'_3, \ell_4$; (c)
Illustration of locations $A_2$ and $B_1$, and paths $\ell'_1,
\ell_2, \ell_3, \ell'_4$; (d) Illustration of locations $A_2$ and
$B_2$, and paths $\ell'_1, \ell'_2, \ell'_3, \ell'_4$. Other
locations and their corresponding paths can be given in a
similar way.} \label{fig2}
\end{figure}

\end{document}